\documentclass[floats,floatfix,showpacs,amssymb,prd,twocolumn,superscriptaddress,nofootinbib,nolongbibliography,reprint]{revtex4-2}

\usepackage{amssymb,amsmath,verbatim,mathtools,needspace,enumitem,etoolbox,graphicx,physics,microtype,afterpage,bigints,gensymb,tabularx,mathtools,siunitx, scalerel}

\usepackage[dvipsnames, usenames]{xcolor}
\definecolor{linkcolor}{rgb}{0.0,0.3,0.5}
\definecolor{dodgerblue}{HTML}{1E90FF}
\usepackage{hyperref}
\hypersetup{unicode, colorlinks=true, linkcolor=linkcolor, citecolor=linkcolor, filecolor=linkcolor,urlcolor=linkcolor, pdfusetitle}
\usepackage[all]{hypcap}
\usepackage[T1]{fontenc}
\usepackage[utf8]{inputenc}
\usepackage{orcidlink}

\usepackage [english]{babel}
\usepackage [autostyle, english = american]{csquotes}
\MakeOuterQuote{"}

\usepackage{soul}
\usepackage{bbm}

\def\aj{{Astron.~J.}}
\def\araa{{Ann.~Rev.~Astron.~Astrophys.}}
\def\apj{{Astrophys.~J.}}\def\apjl{{Astrophys.~J.~Lett.}}
\def\apjs{{Astrophys.~J.~Supp.}}
\def\ao{{Appl.~Opt.}}

\def\aap{{Astron.~Astrophys.}}
\def\aapr{{Astron. Astrophys. Rev.}}

\def\jcap{{J. Cosmology Astropart. Phys.}}

\def\mnras{{Mon.~Not.~R.~Astron.~Soc.}}             

\def\nar{{New A. Rev.}}

\def\prd{{Phys.~Rev.~D}}

\def\pasj{{	Publ. Astron. Soc. Jpn.}}

\def\ssr{{Space~Sci.~Rev.}}

\def\physrep{{Phys.~Rep.}}

\renewcommand{\emph}[1]{\textit{#1}}

\newcommand{\umani}{\affiliation{Department of Physics and Astronomy \& Winnipeg Institute for Theoretical Physics, University of Manitoba, Winnipeg, R3T 2N2, Canada}}

\newcommand{\iiser}{\affiliation{Indian Institute of Science Education and Research, Pune, Maharashtra 411008, India}}

\newcommand{\calg}{\affiliation{Department of Physics \& Astronomy, University of Calgary, Calgary, T2N 1N4, Canada}}

\newcommand{\Msol}{\rm \,M_{\odot}}

\begin{document}

\title{A multi-messenger window into galactic magnetic fields \\ and black hole mergers with LISA}

\author{Anuraag Reddy
$\,$\orcidlink{0009-0008-6390-0678}
}
\iiser

\author{Nathan Steinle$\,$\orcidlink{0000-0003-0658-402X}}
\email{nathan.steinle@umanitoba.ca}
\umani

\author{Samar Safi-Harb$\,$\orcidlink{0000-0001-6189-7665}}
\umani

\author{Jo-Anne Brown$\,$\orcidlink{0000-0003-4781-5701}}
\calg

\begin{abstract}
Large-scale (i.e., $\gtrsim {\rm kpc}$) and micro-Gauss scale magnetic fields have been observed throughout the Milky Way and nearby galaxies. These fields depend on the geometry and matter-energy composition, can display complicated behavior such as direction reversals, and are intimately related to the evolution of the source galaxy. Simultaneously, gravitational-wave astronomy offers a new probe into astrophysical systems, for example the Laser Interferometer Space Antenna (LISA) will observe the mergers of massive (i.e., $M ~> 10^6$ M$_{\odot}$) black-hole binaries and provide extraordinary constraints on the evolution of their galactic hosts. In this work, we show how galactic, large-scale magnetic fields and their electromagnetic signatures are connected with LISA gravitational-wave observations via their common dependence on the massive black-hole binary formation scenario of hierarchical galaxy mergers. Combining existing codes, we astrophysically evolve a population of massive binaries from formation to merger and find that they are detectable by LISA with signal-to-noise ratio $\sim 10^3$ which is correlated with quantities from the progenitors' phase of circumbinary disk migration such as the maximum magnetic field magnitude $|\mathbf{B}| \approx 7 \,\mu$G, polarized intensity, and Faraday rotation measure. Interesting correlations result between these observables arising from their dependencies on the black-hole binary total mass, suggesting a need for further analyses of the full parameter space. We conclude with a discussion on this new multi-messenger window into galactic magnetic fields. 
\end{abstract}

\maketitle

\section{Introduction} 
\label{sec:Intro}

Magnetic fields are present throughout a galaxy and play a vital role in galactic evolution and dynamics \cite{Widrow2002}. Compared to the strong, smaller-scale magnetic fields around supermassive black holes (SMBHs) in galactic cores \cite{EHTm872021,EHTsaga2024}, weak- and larger-scale magnetic fields have been observed in the Milky Way and nearby galaxies \cite{Beck1996,Beck2001,Ferriere2009,VanEck2011,Beck2019,Beck2020,Krause2020,Borlaff2023,Akshaya2024}. Large-scale magnetic fields are thought to originate from seeds such as primordial fields in the early Universe \cite{Totani1999,Widrow2012,Subramanian2019,Martin-Alvarez2021,Pavivcevic2025} and amplified through mechanisms such as the galactic dynamo \cite{Ruzmaikin1988,Beck1996,Kulsrud1999,Widrow2002,Shukurov2021,Brandenburg2023}. These magnetic fields are being actively studied through multi-wavelength electromagnetic (EM) campaigns which are set to be revolutionized by future facilities, such as the Square Kilometer Array  \citep[SKA; e.g.,][]{Gaensler2004,Arshakian2009,Bonafede2015,Heald2020}.
Although the Milky Way and most other galaxies are known to host central SMBHs, there have not been studies that explore how the SMBHs impact the properties of the large-scale magnetic fields present in their host galaxies. We contend that the detection of gravitational waves (GWs) offers new avenues to study galactic magnetic fields and provides a natural connection to SMBHs. 

The Laser Interferometer Space Antenna (LISA) will be a space-based observatory sensitive to milli-Hz GWs that originate from within and beyond the Milky Way \cite{LISA2017}. Among many sources, LISA will indirectly probe the strong magnetic fields of galactic compact binaries \cite{Savalle2024}, the gravito-magnetic field generated by the rotation of the Milky Way's dark matter halo \cite{Tartaglia2021}, and noise contributions due to the interplanetary magnetic field \cite{Cesarini2022,Armano2024}. 
A main science target for LISA are populations of binary mergers of SMBHs, i.e., black holes with masses in the range $\sim 10^5 \Msol$ to $10^9 \Msol$ \cite{LISA2017}. 
Detections of these mergers will provide unprecedented constraints on the progenitor evolutionary pathways and galactic environments \citep[e.g.,][]{LISAastro2023}. 

A canonical example is the formation channel of hierarchical galactic mergers \cite{Colpi2014,Volonteri2021,Pfeifle2024}. In this case, a pair of galaxies, each with its own central SMBH, will merge to form a new galactic environment within which a SMBH binary also forms and inspirals due to orbital angular momentum loss from astrophysical (e.g., stellar, gaseous, and dark matter) processes until GW emission finalizes the SMBH binary merger. 
LISA will only be sensitive to the final moments of this process, necessitating comprehensive models to accurately infer the evolution of the progenitor system \cite{Dotti2012,Dayal2019,Volonteri2020}. 
Simultaneously, hierarchical galaxy mergers are related to the origin and evolution of galactic magnetic fields. 
A three-dimensional hydrodynamical model shows that the galactic merger system can exhibit maxima in the large-scale magnetic field \cite{Rodenbeck2016}, depending on the relative orientations of the merging galactic planes. 
Some studies have reconciled galactic magnetic field models with the physics of hierarchical galaxy mergers, \citep[e.g.,][]{Arshakian2009,Xu2010,Yar-Mukhamedov2015}. 
Modeling of the large-scale magnetic field is sensitive to the mass of the galaxy and to assumptions regarding star formation, supernovae feedback, and ram-pressure stripping \cite{Rodrigues2015}. A corollary is that magnetic field modeling is expected to be important for understanding these smaller-scale processes \cite{Beck2013,Beck2015,Beck2019}.

There is great complementarity between the communities that study galactic magnetic fields and LISA astrophysics rooted in their common study of galaxy mergers. Indeed, an analogous commonality exists regarding interpretations of observations of the cosmic microwave background \citep[e.g.,][]{Grasso2001,Planck2020}. Our aim is to leverage this commonality and reveal the multi-messenger window that LISA offers into the study of galactic magnetic fields. 
For example, LISA will be sensitive to SMBH mergers at high redshift, enabling new studies on galactic magnetism throughout cosmic expansion. 
This new window will also complement multi-messenger strategies that examine coincident EM and GW signals associated with the binary merger \cite{Mangiagli2022,Dong2023}. 

\begin{figure}
\centering
\includegraphics[width=0.48\textwidth]{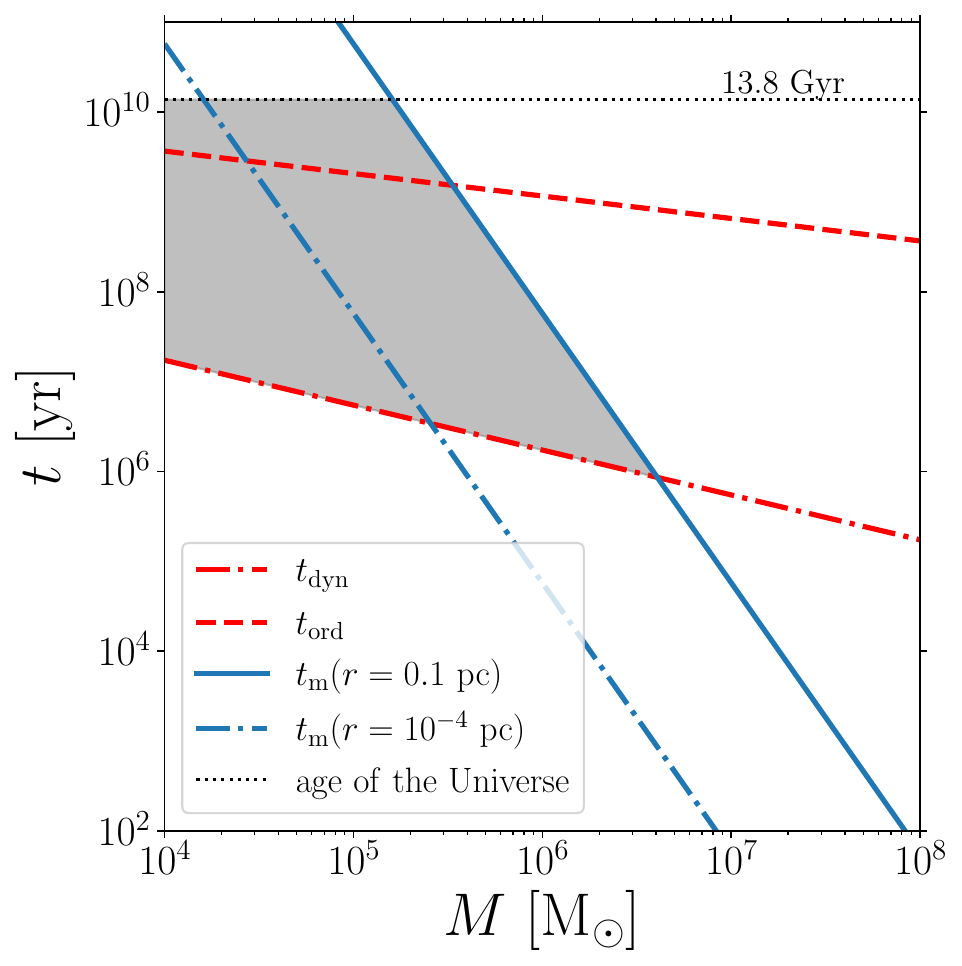}
\caption{
Dependence on the SMBH binary total mass $M$ of the GW-driven SMBH binary merger timescale $t_{\rm m}$ (assuming equal masses), shown by the blue lines for two values of binary separation, and of two timescales relevant for the large-scale galactic magnetic field: 
the dynamo timescale $t_{dyn}$ (dot-dashed red line) 
is the time over which the magnitude of a large-scale field is amplified from the dynamo action, and the ordering timescale $t_{ord}$ (dashed red line) is the time over which the spatial ordering of the field is obtained. 
The gray patch shows an example region of interest: SMBH binaries with $13.8\,{\rm Gyr} > t_{\rm merge} > t_{\rm dyn}$ allowing dynamo amplification in the host galaxy before the merger of the SMBH binary, which is ultimately limited by the age of the Universe (dotted black line) for observable GWs with LISA. 
} \label{F:Timescales}
\end{figure}

Given the co-evolution of the galactic magnetic field and the SMBH binary through the hierarchical galactic merger process, a couple questions naturally arise: how do the processes relevant for large-scale magnetic fields and for SMBH binary mergers overlap in the parameter space? 
And how do we relate the observable messengers of the galactic magnetic field and the SMBH binary merger? 

We sketch an answer to the first question in Figure \ref{F:Timescales} with magnetic field timescales and the GW-driven SMBH binary merger timescale. 
For a differentially rotating thin-disk galaxy, the mean-field dynamo generates a \emph{large-scale} magnetic field from both helical turbulent flows of gas and differential rotation, though the former is responsible for seeding the field and the latter for maintaining the amplification of the field. 
The dynamo timescale $t_{\rm dyn}$ corresponds to the dynamo amplification of a seed magnetic field, and the ordering timescale $t_{\rm ord}$ corresponds to the steady-state time required to reach a coherence length on the order of the galaxy size. 
These can be approximated \cite{Arshakian2009} by $t_{\rm dyn} = h/\Omega l$ and $t_{\rm ord} = (R/l)\left(h/v\Omega\right)^{1/2}$, respectively, where $h$ and $\Omega$ are our SMBH-mass dependent disk thickness and galactic angular velocity, and $l$ and $R$ are the characteristic length and radius of the galaxy, respectively, and $v$ is the turbulent velocity. 
Here we assume $\Omega$ depends on the SMBH mass $M$ as given in Section~\ref{sec:MagField} (see Eq.~(\ref{E:Vdisk2})), and that $h = 0.5$ kpc, $l = 0.05$ kpc, $R = 20$ kpc, and $v = 10$ km/s \cite{Arshakian2009}. 
Generally, $t_{\rm dyn} < t_{\rm ord} \sim 1$ Gyr, implying that the structure of the magnetic field continues to evolve after saturation via amplification before reaching steady state \cite{Arshakian2009}. 
Meanwhile, assuming there is an equal-mass SMBH binary in the core of this galaxy, the timescale for the coalescence \cite{Peters1964} is $t_m \sim 10 \left(r/10^{-4}\,{\rm pc}\right)^{4} \left(M/\num{2e7}\,{\rm M}_\odot\right)^{-3}$~yrs where we assume two values of separation, $r = 0.1$~pc and $r = 0.0001$~pc, typical for a SMBH binary in a galactic circumbinary disk. 

\begin{figure*}
\centering
\includegraphics[width=\textwidth]{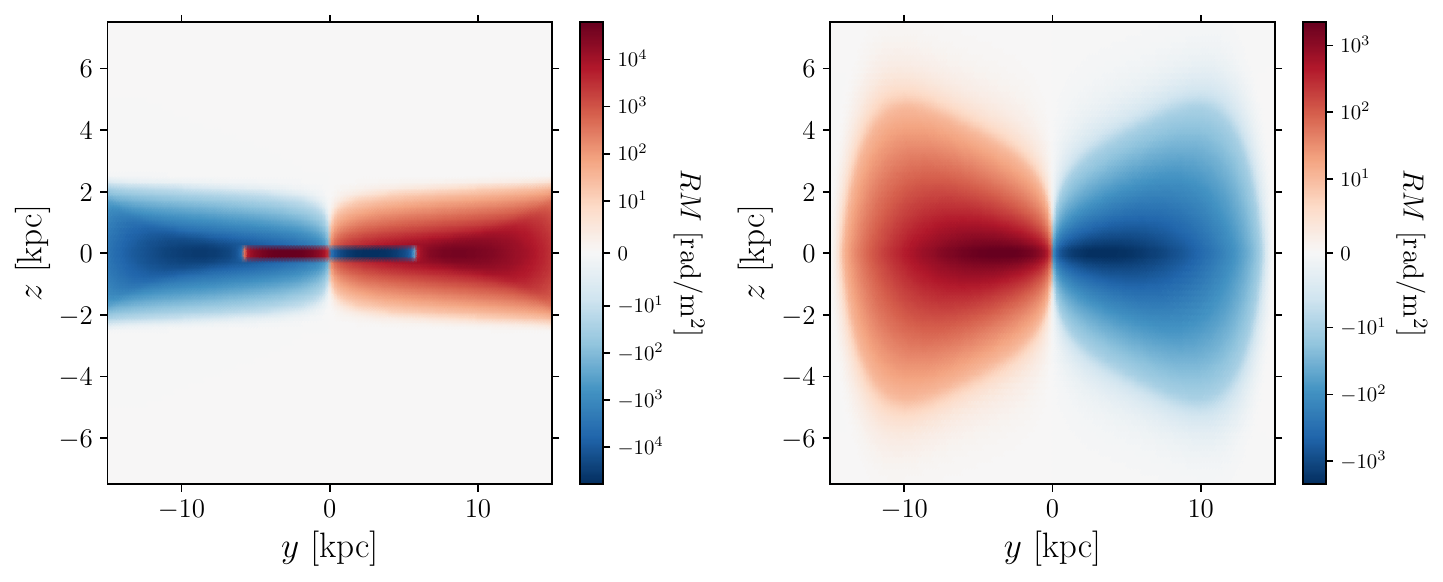}
\caption{The rotation measure ($RM$) computed from \textsc{galmag}, i.e., see Eq.~(\ref{E:RM}), in the $y$--$z$ plane of a galaxy composed only of a thin disk [spherical halo] in the left panel [right panel]. 
We emphasize that this is the rotation of the polarization due to the source galaxy's media, and discuss other possible sources of rotation in Sec~\ref{sec:MultiMess}. In the left panel with the flared disk source, we assume one field reversal at $r = 7$ kpc. 
} \label{F:RM}
\end{figure*}

Combining these in Fig. \ref{F:Timescales} suggests that $M$ is important for understanding whether the magnetic field is amplified and/or spatially ordered prior to the SMBH binary merger, which may be more likely for higher $M$ as $t_{\rm m}$ and $t_{\rm dyn}$ are inversely proportional to $M$. 
More specifically, the relevant range of $M$ over which dynamo-amplified magnetic fields may exist prior to the SMBH merger depends on the binary separation that the SMBH binary finishes the phase of circumbinary disk migration. This is ultimately limited by the need for SMBH binaries that merge within the age of the Universe, shown by the dotted black line. Fig. \ref{F:Timescales} suggests that (i) GW observations can probe galactic magnetic fields as these processes overlap in the parameter space, and (ii) the large-scale magnetic field strength at a progenitor stage of the SMBH binary merger could be sufficient to enable multi-messenger studies of galactic magnetic fields. 

To demonstrate this complementarity and illustrate its implications, we employ the semi-analytic model of hierarchical SMBH binary evolution in \texttt{holodeck} \cite{holodeck} to evolve each system, generated from dark-matter halo merger trees of the Illustris cosmological simulation, through phases of dynamical friction, stellar loss-cone scattering, and circumbinary disk evolution before GW emission finishes the SMBH binary coalescence. To connect this with the physics of galactic magnetic fields, we utilize \textsc{galmag} \cite{Shukurov2019}, which analytically solves the mean-field dynamo equation, to compute the large-scale magnetic field and EM observables, such as the polarized intensity and Faraday rotation measure, in the circumbinary disk phase. 
Importantly, we modify \textsc{galmag} to include dependence of the magnetic field on the total mass of the binary SMBH. 
Then we compute the LISA response to each SMBH binary merger via time-delay interferometry with \texttt{balrog} \cite{balrog1} and analyze its correlations with the magnetic field properties and with EM observables from the progenitor's circumbinary phase.  
Our analysis reveals interesting correlations in this new multi-messenger parameter space and motivates future studies with improved modeling techniques. 

This paper is organized as follows. We present the adopted model of the galactic magnetic field in Sec.~\ref{sec:MagField}, which we apply to a population of SMBH binaries in Sec.~\ref{sec:LISA}. In Sec.~\ref{sec:MultiMess} we discuss multi-messenger implications of our results, and we summarize our conclusions in Sec.~\ref{sec:ConcDisc}.

\section{Model of the large-scale galactic magnetic field}
\label{sec:MagField}

Galactic magnetic fields have been studied for nearly a century \cite{vandeHulst1967} with primary focus on the Milky Way and nearby galaxies \cite[e.g.,][]{Beck2001,Beck2009,Beck2019}.  
Common techniques for probing large-scale magnetic fields include the analysis of the polarization properties of synchrotron and thermal emission from compact sources and cosmic rays (high-energy electrons spiraling around magnetic field lines), and the Faraday rotation of polarized light propagating through magnetized plasma \citep[e.g.,][]{BrownEtAl:2007, VanEck2011, OrdogEtAl:2017,Hutschenreuter2022,Dickey2022}.

Dynamo theory is a common mechanism used in the modeling of such magnetic fields, and there exist many numerical and semi-analytic approaches to solving the mean-field dynamo equation, 
\begin{align}\label{E:Dynamo}
\begin{aligned}
    \frac{\partial \mathbf{B}}{\partial t} = \nabla \times (\alpha \mathbf{B}) + \nabla \times (\mathbf{V} \times \mathbf{B}) + \beta \nabla^2 \mathbf{B}
\end{aligned}
\end{align}
where $\mathbf{B}$ is the large-scale magnetic field, $\alpha$ and $\beta$ are the coefficients representing the $\alpha$-effect and the turbulent magnetic diffusion effect, and $V$ is the large-scale velocity field 
arising mostly from differential rotation. 
Eq~(\ref{E:Dynamo}) encapsulates the large-scale magnetic field's dependence on the random flows arising from the differential rotation and helical turbulent motions of gas. 

Models such as the mean-field dynamo can be constrained by observations of the galaxy's EM emission \citep[e.g.,][]{Beck2015,Krause2020}, and are explored theoretically with simulations of galaxy evolution and structure \cite[e.g.,][]{Rodenbeck2016,Pakmor2017}, We use \textsc{galmag} \cite{Shukurov2019} to model the galactic magnetic field. \textsc{galmag} efficiently parameterizes solutions to Eq~(\ref{E:Dynamo}) via an expansion in free-decay modes for spherical and disk geometries, which provides a linearized approximation to the full non-linear galactic dynamo dynamics. 
\textsc{galmag} was developed with purpose to help provide an accessible model that mathematically covers a diversity of magnetic field configurations, allowing computation of the large-scale magnetic field $\mathbf{B}$ in three-dimensional space. 

The \textsc{galmag} model depends on free parameters that relate the dynamo to the properties of the galaxy. Throughout this work, we assume the default values of these parameters as given in Table 1 of \cite{Shukurov2019}, most important for the present analysis are the radii for the disk and halo that control the active dynamo region and the shear rate due to differential rotation and the intensity of helical turbulence. 
We emphasize that a realistic model would obtain the values of these parameters from a cosmological simulation that tracks the properties of the galaxies. We discuss this further in Sec.~\ref{sec:ConcDisc}. 

We can then calculate the Stokes parameters $I, Q$, and $U$ of the synchrotron emission as described in Appendix C of \cite{Shukurov2019}, which depend on the wavelength $\lambda$, the component of the magnetic field along the line of sight, and the electronic cosmic ray distribution (which we assume to be uniform) of the source. Taking a frame where the $x$-direction is along the line of sight, $I, Q$, and $U$ are functions of the $y$ and $z$ coordinates. Given a value of $\lambda$, we can construct EM observables such as the linear polarized intensity $PI = \sqrt{Q^2 + U^2}$, an example of which from \textsc{galmag} is provided in Figure 13 of \cite{Shukurov2019}. Faraday rotation describes the angular evolution of linearly polarized radiation. The rotation measure ($RM$) is the measurable consequence of Faraday rotation due to the source galaxy, and is given by 
\begin{equation}\label{E:RM}
    RM = \frac{\partial \Psi}{\partial \lambda^2} = 0.81\ \text{rad} \int \frac{n_e(x',y,z)}{1\ \rm cm^{-3}} \frac{B_x(x',y,z)}{1\ \rm \mu G} \frac{dx'}{1\ \rm pc}\, .
\end{equation}
Here, $B_x$ is the $x$-component of the galactic magnetic field, $n_e$ is the thermal electron density (i.e., Eq. (69) of \cite{Shukurov2019}), and $\Psi = \frac{1}{2}\arctan(U/Q)$ is the observed polarization angle. 
There can be many additional sources of rotation along the line of sight between the source galaxy and the observer, including the Milky Way and intergalactic media, which we revisit in Sec.~\ref{sec:MultiMess}. 
We note that the sign of the integral in Eq.~(\ref{E:RM}) as implemented in \textsc{galmag} is contained in $B_x$. 

\begin{figure}
\centering
\includegraphics[width=0.48\textwidth]{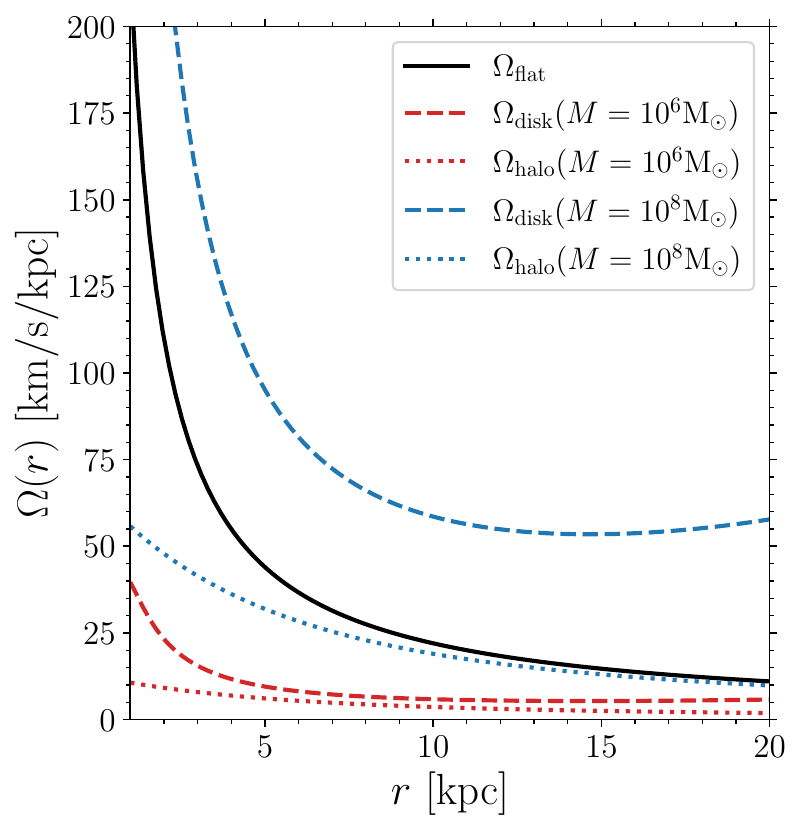}
\caption{Angular velocity profiles versus the galactic radius $r$. 
The solid black line is the default profile for a rigidly rotating disk. 
Our SMBH-mass dependent profiles for the spherical halo in dotted lines (i.e., Eq.~(\ref{E:vel_gas})) and the thin disk in dashed lines (i.e., Eq.~(\ref{E:Vdisk2})) are shown for two values of the SMBH mass, $M = 10^6 \Msol$ in red and $M = 10^8 \Msol$ in blue. We assume the SMBH binary resides in the (unresolved) galactic center.  
} \label{F:VelocityShear-massdep}
\end{figure}

\begin{figure}
\centering
\includegraphics[width=0.48\textwidth]{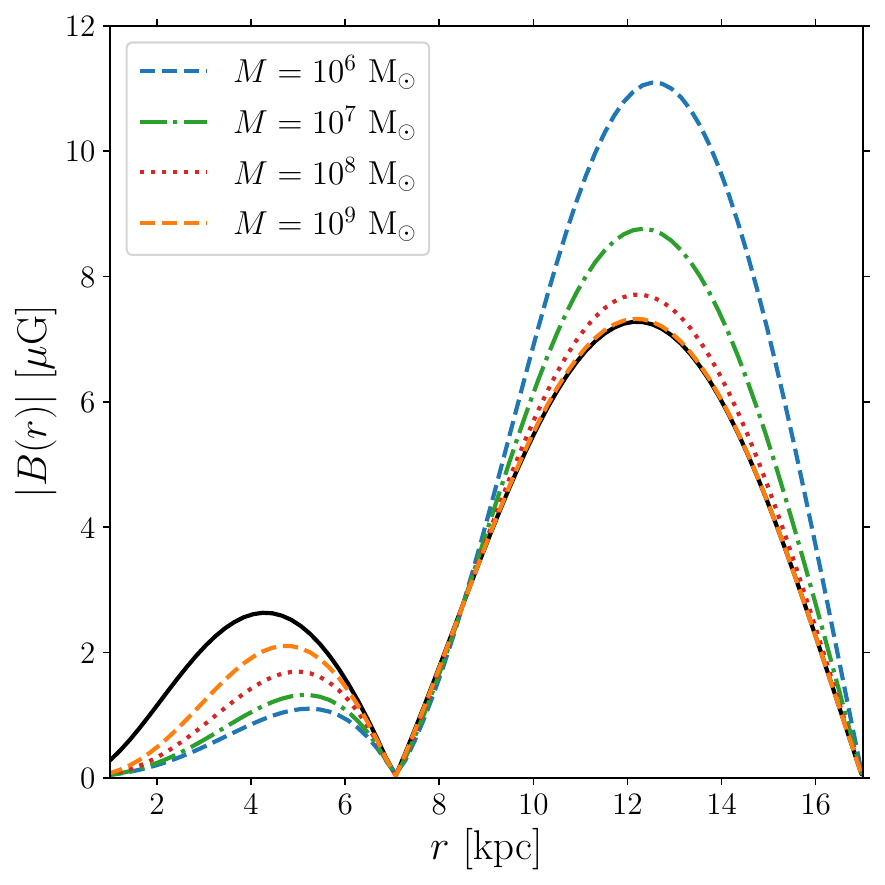}
\caption{
The dependence of the magnitude of the galactic magnetic field in the galactic plane as a function of galactic radius assuming one field reversal at $r = 7$ kpc. 
The black solid line is the mass-independent output of \textsc{galmag}, i.e. using the default velocity profiles, whereas the dashed lines result from our SMBH mass-dependent velocity profiles for $M = 10^6,\ 10^7,\ 10^8$ and $10^9 \Msol$ shown by dashed blue, red, orange, and cyan lines, respectively. 
Here (and in the remainder of this article) we assume the galaxy is composed of both thin disk and spherical halo sources of the magnetic field.  
}\label{F:Bfieldmag}
\end{figure}

To compute the magnetic field, we use a Cartesian coordinate grid, and we assume the default settings of \textsc{galmag} for solutions to Eq.~(\ref{E:Dynamo}). 
Figure \ref{F:RM} illustrates generic output of the \textsc{galmag} galactic magnetic field model with synthetic maps of the $RM$, Eq.~(\ref{E:RM}), in the $y-z$ plane of a galaxy composed only of a thin disk (spherical halo) in the left (right) panel. 
In this edge-on view of the galaxy, the bilateral symmetry across the $y = 0$ axis in both panels is due to the geometrical symmetry of the thin disk and spherical halo sources. In the left panel, the features in $RM$ that extend to $|z| \sim 2$ kpc correspond to the flared disk scale height (i.e., the alternative of a constant disk height produces a $RM$ profile that is only nonzero between $|z| \lesssim h_0 = 0.5$ kpc where $h_0$ is the assumed reference scale height). The component of the magnetic field along the line of sight $B_x$ crosses zero at $y \sim \pm 6$ kpc and $|z| \sim \pm h_0 = 0.1$ kpc causing an unphysical, steep sign change in $RM$ at these locations (we confirmed this occurs for the case of a constant disk scale height as well). 
The maximum and minimum values of $RM$ are larger for the disk source than for the halo source as the magnitude of the galactic magnetic field is larger from the disk than the halo. 
The large value $|RM| \sim 10^5$ rad/m$^2$ near the galactic center is consistent with EM observations of the Milky Way's galactic center \citep[e.g.,][]{EHTsaga2024, wielgus2024} and the pattern bears the iconic X shape of the emission. 
For the thin, flared disk we assume a magnetic field reversal at $r = 7$ kpc, causing the colors in the left panel and right panels to be reversed in Fig.~\ref{F:RM}, i.e., blue $\leftrightarrow$ red, as the field reversal amounts to a reflection across the x-axis and a sign change of $B_x$. 

An important free parameter in \textsc{galmag} is the specification of reversals in the azimuthal direction of the magnetic field at chosen radii. 
Magnetic field reversals have been observed and studied via EM methods in the Milky Way for decades \cite[e.g.,][]{SK1980, Vallee1996, HanEtAl2006, BrownEtAl:2007, OrdogEtAl:2017, Curtin2024}, though the number and exact location of them remains hotly debated.  Interestingly, only one other spiral galaxy (NGC4666) has been observed to have a large-scale reversal along the radial direction \cite{Stein2019}. 
The existence of large-scale reversals have important implications for the total strength of the magnetic field. 
In the \textsc{galmag} framework the field reversals are treated as boundary conditions for the mode numbers in the solutions to Eq.~(\ref{E:Dynamo}). 
We find generally that, all else being equal, the maximum magnetic field magnitude increases with the mode number and also the number of modes. 

An important feature of \textsc{galmag} is the linearity of its solutions that allow one to construct the magnetic field from the disc and the halo separately. As such, we input rotation curves corresponding to each component, which crucially allows us to input curves that are different from the default rotation curves. We utilize this design to connect the galactic magnetic field to the SMBH binary mass $M$ by constructing rotation curves that have a dependence on the SMBH binary at the core of the galaxy. We take a different approach in modifying the rotation curve for each source component. These modifications do not alter the underlying assumptions of \textsc{galmag} as we essentially insert a new, SMBH-mass dependent velocity profile. 
Throughout this article, the galaxies we consider are composed of both thin disk and spherical halo sources unless otherwise specified, which can produce complicated magnetic field structure, e.g. see Fig. 11 of \cite{Shukurov2019}. 

For the gaseous halo source, \textsc{galmag} has a default velocity profile defined as
\begin{align}\label{E:vel_gas}
\begin{aligned}
    \mathbf{v}(r, \theta) = V_h \frac{1-e^{-s/s_v}}{1-e^{-r_h/s_v}}  \boldsymbol{\skew{-.5}\hat\phi}  \,,  
\end{aligned}
\end{align}
where $s = r\sin\theta$, $r_h = 15$ kpc is the radius that defines the active dynamo region of the halo, and $s_v = 3$ kpc is a turnover radius of Milky Way-type galaxies.  
To implement dependence of the binary SMBH total mass $M$ in the gaseous halo velocity profile, we use a standard $M$-$\sigma$ relation derived from EM observations of single SMBHs \cite{McConnell2013}, 
\begin{align}
\begin{aligned}\label{E:Msigma}
    \frac{M}{10^8 \Msol} = \left( \frac{\sigma_{\rm *}}{200\ \rm km/s}\right)^{5.6} \,. 
\end{aligned}
\end{align}
This can be related to the gaseous halo velocity dispersion by \cite{sami_gas_stellar_rel}, 
\begin{align}
\begin{aligned}\label{E:SAMI}
    \log\sigma_{\rm *} = 0.59\log\sigma_{\rm g} + 0.91\,,  
\end{aligned}
\end{align}
where we define the velocity dispersion of the gas as 
\begin{align}
\begin{aligned}\label{E:sigma2}
    \sigma_{\rm g}^2 = 2\pi \left(\frac{4}{3}\pi r_h^3 \right)^{-1} \int_{0}^\pi\int_{0}^{r_h}(\scaleleftright{<}{v}{>} - v(r,\theta))^2r^2\sin\theta dr d\theta\,, 
\end{aligned}
\end{align}
and the average of $v = |\mathbf{v}(r, \theta)|$ is
\begin{align}\label{E:mean_v}
\begin{aligned}
    \scaleleftright{<}{v}{>} = 2\pi \left(\frac{4}{3}\pi r_h^3 \right)^{-1} \int_{0}^\pi \int_{0}^{r_h}v(r,\theta)r^2\sin\theta dr d\theta\,, 
\end{aligned}
\end{align}
Substituting Eq (\ref{E:vel_gas}) into Eq.'s (\ref{E:sigma2}) and (\ref{E:mean_v}) reveals that $\sigma_{\rm g} = ({\rm constant}) V_h $, as $V_h$ is independent of $r$ and $\theta$. To calibrate the velocity profile to $M$, we scale the velocity profile in Eq (\ref{E:vel_gas}) with the mass-dependent factor $V_h = \sigma_{\rm g}/{\rm constant}$. 

For the thin disk source, \textsc{galmag} assumes by default an exponential scale height $h(r)$, i.e., the thickness of the disk increases with radius, and an exponential velocity profile. We choose to keep the same scale height functional dependence (i.e., a flared disk) and to modify the velocity profile. The modified velocity profile is obtained from the velocity dispersion $\sigma_{\rm d}$ of the disk,  
\begin{align}\label{E:Vdisk1}
\begin{aligned}
    v_{\rm d}(r) \approx \left( R_f\frac{d}{dr}[\sigma_{\rm d}^2(r)] \right)^{1/2}  \,,  
\end{aligned}
\end{align}
where we set $R_f = 50$ kpc. We assume $\sigma_{\rm d}$ depends on the (local) mass surface density $\Sigma(r)$ as \cite{Sharma2014}, 
\begin{align}\label{E:Sigmadisk1}
\begin{aligned}
    \sigma^2_{\rm d}(r) =  G h(r) \Sigma(r) \,,
\end{aligned}
\end{align}
where $G$ is the gravitational constant. 
Using the self-gravitation criterion of the disk at its fragmentation radius \citep[e.g.,][and references therein]{Gerosa2015} , one can approximate $\Sigma(r)$ as 
\begin{align}\label{E:MassDensity}
\begin{aligned}
    \Sigma(r) =  \frac{M_{\rm d}(r)}{4\pi r^2} = \frac{M}{4\pi r^2}q_{\rm d}(r) \simeq \frac{M/{10^7 \Msol}}{4\pi (r/{\rm kpc})^2} \left(\frac{r}{\rm kpc} \right)^{\delta}   \,,
\end{aligned}
\end{align}
where $q_{\rm d}(r)$ is the ratio of the disk mass $M_{\rm d}(r)$ to the black hole binary total mass $M$, and $\delta$ is a spectral index. Substituting Eq.~(\ref{E:MassDensity}) into Eq.'s~(\ref{E:Sigmadisk1}) and (\ref{E:Vdisk1}), one obtains 
\begin{align}
\begin{aligned}
    \sigma^2_{\rm d}(r) =  \frac{43}{4\pi}   \frac{h_0}{\rm kpc}\frac{M}{10^7 \Msol}  e^{(r - r_0)/r_d} \left(\frac{r}{\rm kpc}\right)^{\delta - 2} \left(\frac{\rm km}{\rm s}\right)^{2} \, 
\end{aligned}
\end{align}
and 
\begin{align}\label{E:Vdisk2}
    v_{\rm d}(r) &= v_0 \left(\frac{43}{4\pi}\frac{h_0}{\rm kpc} \frac{M}{10^7 \Msol} \frac{R_f}{r_d} e^{(r - r_0)/r_d} \left(\frac{r}{\rm kpc}\right)^{\delta - 2} \right)^{1/2} \notag \\
    & \times \left(1 + \frac{1}{2}(\delta - 2)\left(\frac{r}{\rm kpc} \right) \right)\, \frac{\rm km}{\rm s} \,.
\end{align}
respectively, where $v_0 = 220$ km/s is the scaling consistent with \textsc{galmag}, $h_0 = 0.5$ kpc is the assumed reference scale height, $r_0 = 8.5$ kpc is the assumed reference galactocentric radius, and $r_d = 17$ kpc is the radius that controls the active dynamo region of the disk. 
We then compute the corresponding shear rate $S = rd\Omega/dr$ to ensure compatibility with \textsc{galmag} for angular velocity $\Omega = v_{\rm d}(r)/r$. 
Throughout this work we set $\delta = 4$ to ensure a sufficient correlation with $M$ and convergence of the $\textsc{galmag}$ solution. 
While this differs from standard semi-analytic viscous thin disks, where e.g. $\delta = 1/20$ \cite{Gerosa2015}, but this choice has broad implications that require extended analysis. 

First, Figure \ref{F:VelocityShear-massdep} shows the mass-dependent angular velocity $\Omega$ for the gaseous halo (dotted lines) and thin disk (dashed lines) sources of the magnetic field assuming two values of SMBH binary mass $M = 10^6 \Msol$ (red lines) and $M = 10^8 \Msol$ (blue lines). The exponential form of the profile is preserved for both sources except for the thin disk in the case of large mass where the radial dependence dominates at large galactic radii. 
As the dependence of $\Omega$ on $M$ in our model is ad-hoc it is not intended to be a robust prediction for astrophysical properties of galaxies. Instead, it demonstrates the plausibility of such dependence in order to reveal important consequences for future EM and GW observations. 
Nonetheless, the existence of such a dependence between $M$ and $\Omega$ is known from studies of gas kinematics around SMBHs, a common observational method for determining $M$ \cite{Kormendy2013}, but the precise dependence of $\Omega$ on $M$ and the resulting correlation with the magnetic field is uncertain \citep[e.g.,][]{VerdoesKleijn2002}. 

\begin{figure}
\centering
\includegraphics[width=0.48\textwidth]{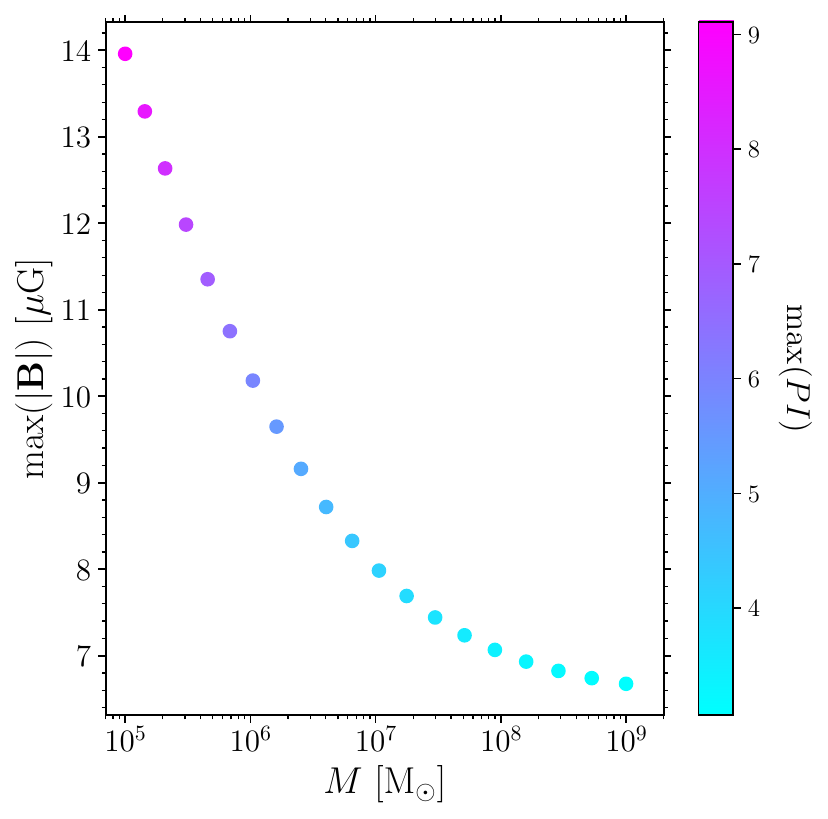}
\caption{
An example of the correlations between the SMBH binary total mass $M$, the maximum magnitude of the total galactic magnetic field $\mathbf{B}$, and the maximum of the $\lambda = 5$ cm polarized intensity $PI$ in the colorbar.  
} \label{F:Scatter}
\end{figure}

Next, Figure \ref{F:Bfieldmag} depicts how the radial dependence of the magnitude of the magnetic field in the galactic plane $|\mathbf{B}(r)|$ varies with the total mass $M$ of the central SMBH binary. 
The black solid line corresponds to the case of no mass dependence, i.e., the default velocity profiles, and the color non-solid lines result from our binary SMBH mass-modified model for a few values of $M$. 
Generally, the magnetic field vanishes at large radii, i.e. $\gtrsim 16$ kpc, due to the two \textsc{galmag} free parameters that define the active dynamo regions, $r_h$ and $r_d$. 
The height, number, and radial locations of the maxima arise from various aspects of the magnetic field model.  
As discussed earlier, the radial locations of field reversals are free parameters in $\textsc{galmag}$ and Fig. \ref{F:Bfieldmag} assumes one reversal at $r = 7$ kpc. 
Focusing on the black solid line, we see that the number and locations of field reversals affects the maximum magnetic field magnitude, as it shifts the dominant modes of the mean-field dynamo solutions, but also affects the total number of magnetic field maxima and their locations. This is consistent with Figure 7 of \cite{Shukurov2019} where a flared disk with one field reversal produces a larger magnitude at large radii compared to smaller radii, and differences with our Fig. \ref{F:Bfieldmag} are due to our use of both thin disk and halo whereas Fig. 7 of \cite{Shukurov2019} only assumes a thin disk source. 
The dependence on the SMBH binary mass $M$, i.e. the dashed lines in Fig.~\ref{F:Bfieldmag}, is consistent with Fig.~\ref{F:VelocityShear-massdep}, i.e. the order of the dashed lines switches across the field reversal at $r = 7$ kpc due to the velocity profile also reversing direction. 

Figure \ref{F:Scatter} shows the maximum magnetic field magnitude ${\rm max}(|\mathbf{B}|)$ and the maximum polarized intensity ${\rm max}(PI)$ as the binary mass SMBH $M$ increases over a range relevant for the LISA detection horizon. 
The monotonic dependence of the maximum magnetic field on $M$ translates to monotonicity in ${\rm max}(PI)$. 
Consistent with the left panel of Fig.~\ref{F:Bfieldmag}, larger $M$ yields smaller max($\mathbf{B}$) and ${\rm max}(PI)$ which depends directly (but nontrivially) on the sum $B_{\rm y}^2 + B_{\rm z}^2$. We emphasize that future studies will be needed to uncover the nature of these dependencies.  

We have demonstrated how our model translates the central SMBH binary's mass dependence of the galactic magnetic field into the corresponding EM observables. In the next section, we will apply this framework to a population of merging SMBH binaries. 

\begin{figure*}
\centering
\includegraphics[width=\textwidth]{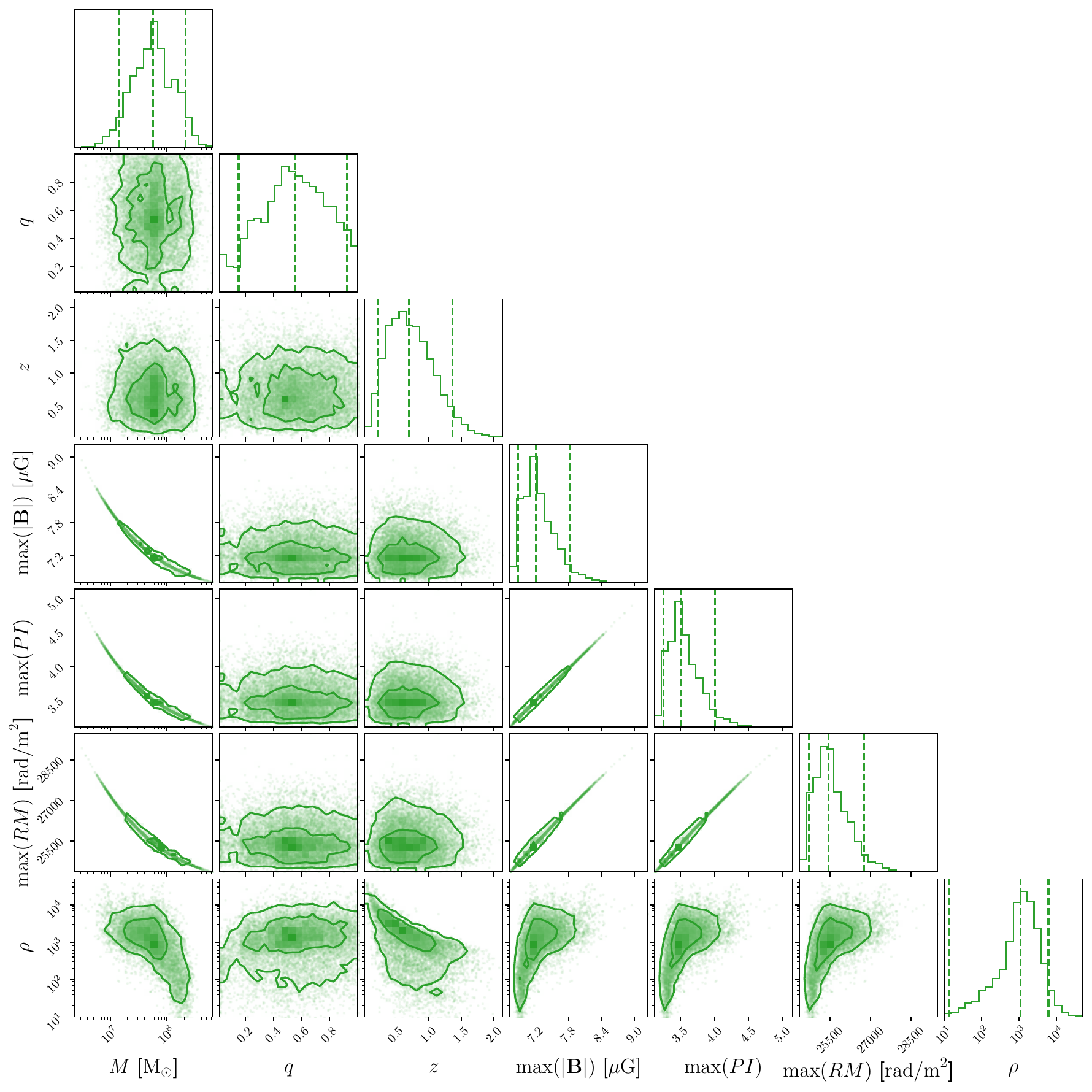}
\caption{Correlations between the initial parameters of the cosmological population of SMBH binaries, the total mass $M = m_1 + m_2$, mass ratio $q = m_2/m_1$, and redshift $z$, and the LISA signal-to-noise ratio $\rho$ of the mergers, and the maximum of the large-scale magnetic field ${\rm max}(|\mathbf{B}|)$, the polarized intensity ${\rm max}(PI)$ at wavelength $\lambda = 5$ cm, and the rotation measure ${\rm max}(RM)$ in the region of each galaxy subtended by 1 kpc $\lesssim y \lesssim$ 10 kpc and 1 kpc $\lesssim z \lesssim$ 10 kpc. 
} \label{F:Corner}
\end{figure*}

\section{LISA detections and magnetic fields of a binary black hole population}
\label{sec:LISA}

In this section, we embed the model developed in the previous section into the broad context of hierarchical galactic formation. This is generally a challenging task, and evolving the galactic magnetic field and SMBH binary in time requires self-consistent and simultaneous simulation of galactic mergers on ${\rm kpc}$ and sub-${\rm kpc}$ scales. 
As this is the first study to attempt this, we compute the magnetic field at a single instant of the hierarchical merger process and assume each galaxy is like the Milky Way. 

We utilize the \texttt{holodeck} \cite{holodeck} code to (i) access a statistical population of inspiraling binary SMBHs generated from the Illustris cosmological simulation \cite{Nelson2015,Kelley2017} and (ii) evolve the population through key astrophysical binary we've processes of the hierarchical formation channel via semi-analytic prescriptions. Before gravitational-wave emission dominates the SMBH binary inspiral, the three classic hardening processes are dynamical friction, scattering of individual stars through the binary's loss-cone, and circumbinary disk (CBD) migration. 

\begin{table*}
\caption{\label{Tab:CodesSummary}
Descriptions of the frameworks used in this study to connect binary SMBH population astrophysics with the properties of the large-scale magnetic field of the galactic host and with the LISA observability of the resultant binary SMBH mergers.  
}
\vspace{0.1cm}
\def\arraystretch{1.2}  % for the vertical padding
\centering
\begin{tabular}{c|c|c}
  & Description & Ref. \\
  \hline
  \texttt{holodeck} &   astrophysical evolution of Illustris binary SMBH population & NANOGrav Collaboration \cite{holodeck} \\
  \textsc{galmag}  &  galactic magnetic fields; modified here to include SMBH mass dependence & \cite{Shukurov2019} \\
  \texttt{balrog} &  multi-purpose data analysis suite for the Laser Interferometer Space Antenna & U. of Birmingham \cite{balrog1,balrog2} \\
  \hline
\end{tabular}
\end{table*}

This approach provides a realistic population and strongly-motivated astrophysical BH-binary evolution. However, it is not self-consistent as \texttt{holodeck} currently does not provide information regarding each galaxy's morphology and \textsc{galmag} was intended for axially symmetric galaxies, and the model of CBD evolution implemented in \texttt{holodeck}, i.e. \cite{Siwek2023}, does not provide a velocity profile to connect it to the large-scale magnetic field (i.e., we instead use the model described in Sec.~\ref{sec:MagField}). These are important uncertainties that we leave for future studies to explore. 
We use the same $M-\sigma$ relation, Eq.~(\ref{E:Msigma}), to introduce mass dependence in the gaseous halo's magnetic field in \textsc{galmag} as is used in \texttt{holodeck} to convert the galactic subhalo parameters of the Illustris simulation into SMBH binary masses. 

Specifically, via \texttt{holodeck} we initialize the Illustris population, evolve it through dynamical friction and binary loss-cone stellar scattering phases and until the end of the CBD phase. At this point, we obtain the SMBH binary total mass $M$ to compute the large-scale magnetic field and EM observables, assuming all other properties are that of the Milky Way, of each galaxy in the population. 
To reduce computation time the modified \textsc{galmag} code is run in parallel. 
We finish the SMBH binary evolution of each system under GW emission until merger in the LISA band. 

To model the LISA detection of the GW's from the SMBH binary coalescence, we use \texttt{balrog} \cite{balrog1,balrog2} which implements time-delay interferometry \cite{Tinto2021} to compute the source's signal-to-noise ratio (SNR) $\rho$ and parameter uncertainties. We use a phenomenological approximant for the full inspiral, merger, and ringdown GW waveform model   \textsc{IMRPhenomXHM}~\citep{Pratten2020,GarciaQuiros2020} which is calibrated to numerical relativity simulations and assumes the spins of the SMBHs are aligned with the binary orbital angular momentum. We utilize higher modes (i.e., GW angular multi-pole modes $\ell \geq 2$) and assume an instrumental lower-frequency cut-off of $10^{-4}$ Hz. Generally, the LISA SNR $\rho$ of merging SMBH binaries has complicated dependence on the SMBH mass and redshift, peaking in $M \sim 10^6 - 10^7$ for $z \lesssim 10$ and vanishing quickly for $M \gtrsim 10^8$ \cite{LISA2017}. 

Combining these ingredients provides a view into the multi-messenger correlations of the source observables. 
Figure \ref{F:Corner} displays the distributions of parameters that result for this population: the total mass $M = m_1 + m_2$; mass ratio $q = m_2/m_1$;  cosmological redshift, $z$, of the SMBH binaries; the maximum of the large-scale magnetic field, ${\rm max}(|\mathbf{B}|)$, across the galaxy; the maximum polarized intensity, ${\rm max}(PI)$, at $\lambda = 5$ cm across each source galaxy;  the maximum rotation measure, ${\rm max}(RM)$, in the region of each galaxy subtended by 1 kpc $\lesssim y \lesssim$ 10 kpc and 1 kpc $\lesssim z \lesssim$ 10 kpc (i.e., a rectangular region in the upper-right quadrant of the panels in Fig.~\ref{F:RM}) at the end of the circumbinary phase (i.e., the beginning of the GW phase); and the SNR $\rho$ of the SMBH mergers as seen by LISA. 

Several interesting correlations are seen in the panels of Fig. \ref{F:Corner}. The three SMBH binary parameters, $M$, $q$, and $z$, are needed to compute the GW waveform and hence LISA SNR $\rho$ whose median value $\sim 10^3$ would enable precise measurements of the SMBH parameters and thus provide unprecedented constraints on the progenitor evolution of the binaries. 
The cross-correlation of $\rho$ with $M$ and $z$ is consistent with astrophysical expectations for the LISA detector \cite{LISA2017}, where moderate $M$ and low $z$ produce the highest $\rho$, and has a tail toward $\rho \lesssim 100$ composed of systems with $M \gtrsim \num{2e8}\Msol$.  

Meanwhile, the maximum magnetic field strength, ${\rm max}(|\mathbf{B}|)$, decreases monotonically as $M$ increases across the population, because we find numerically that ${\rm max}(|\mathbf{B}|) \sim 1/\Omega(r_{\rm max})$ where $r_{\rm max}$ is the radius at which $\mathbf{B}$ is maximum and the galactic angular velocity $\Omega \sim M^{1/2}$ via Eq.~(\ref{E:Vdisk2}). This translates into the EM observables $PI$ and $RM$. 
Recall that we compute these at the end of the phase of circumbinary evolution which precedes the final evolutionary phase of GW-driven inspiral and merger. 
As the total mass $M$ varies across the population, ${\rm max}(PI)$ decreases monotonically consistent with Fig.~\ref{F:Scatter}, and ${\rm max}(RM)$ increases in the specified region of the source galaxy consistent with Fig.~\ref{F:RM}.  
Together, these produce complicated correlations. For example, the panel in the the first row and fourth column shows how ${\rm max}(RM)$ is anti-correlated with the LISA SNR due to their joint dependence on the SMBH mass, $M$, and the fact that $\rho$ vanishes for $M \gtrsim 10^8\Msol$. This correlation depends on the region of interest and hence on the geometrical morphologies of the source galaxies. 

We find that $B_{\rm max}$ occurs at a radius $r_{\rm max} \approx 12$ kpc which is insensitive to $M$, as suggested by Fig.~\ref{F:Bfieldmag}. 
The direction of $\mathbf{B}$ is also independent of $M$ in our model except near the locations of reversals where the polar angle of $\mathbf{B}$ diverges. 
In an extended analysis, one would be interested in probing the \emph{number} of maxima in B, rather than taking the global maximum of B as done here, as this is a proxy for the number of reversals, i.e. there are $n-1$ reversals for $n$ maxima in an idealized case. Some caveats of our analysis are important to discuss. 
We do not self-consistently connect the galactic host environment of the SMBH binary with its galactic magnetic field, as was done in e.g. \cite{Rodrigues2019}. Effectively, our model assumes each galaxy in the population has Milky-Way type properties. This limiting aspect of our work is connected to another uncertainty in our model: the galaxy morphology. 
Although it is thought that the galactic dynamo is not effective at explaining the magnetic field of non-spiral galaxies \citep[e.g.,][]{Widrow2002}, we assume all the SMBH binaries used in Section~\ref{sec:LISA} have axial symmetry, i.e. are spiral, when computing the magnetic field with \textsc{galmag} despite the fact these binaries come from a mixed morphology galactic population \cite{Nelson2015}. 
We do not compute the velocity of the gas in the gaseous halo and disk from their combined gravitational potential with stars and a central SMBH binary, as is standard in studies of the environments of SMBHs, \citep[e.g.,][]{Sofue2013}. 
We also do not time evolve the galactic magnetic field along with the galactic source evolution, as will be necessary for realistic models of the magnetic field's dependence on the SMBH binary mass, which may increase due to accretion, and for precise predictions of future observatories. 
Our work indicates a need for more studies that explore the complicated correlations between the velocity profiles of the sources of the galactic magnetic field, processes related to the central SMBH, and their observational signatures. 

The magnetic field of the galactic host at the end of the circumbinary phase of evolution depends only on $M$ in our model, which is constant under the subsequent GW-driven inspiral and merger. Our assumption that the magnetic field will be sufficiently seeded and dynamo-amplified in the post- galactic merger environment is consistent with the results of studies of the magnetic field's redshift evolution \citep[e.g.,][]{Arshakian2009,Rodrigues2019}, which show that the maximum magnitude of the magnetic field increases through cosmic time and saturates at $\gtrsim 1\ \mu$G for redshifts $z \lesssim 2$ but they do not consider the central SMBHs. 
Ultimately, our results imply a great potential for future studies that simultaneously model the evolution of a galaxy's large-scale magnetic field and its central SMBH binary as possible multi-messenger sources.

\section{Multi-messenger implications}
\label{sec:MultiMess}

In the results of the previous section, we demonstrated how to connect LISA GW observations of merging SMBHs with the properties and EM signatures of large-scale magnetic fields of the progenitor galactic host. 
This implies that LISA data analysis and astrophysical modeling can predict the magnetic field properties and EM signatures of the progenitor of a SMBH merger, analogous to other astrophysics relevant for LISA sources. 

Furthermore, our results propound numerous implications for multi-messenger studies of galactic magnetic fields where non-coincident EM and GW observations provide independent probes of the \emph{population} of large-scale galactic magnetic fields. 
For example, this motivates population analysis to jointly reconcile EM and GW observations with the astrophysics of magnetic fields in hierarchical galaxy mergers.  
It also motivates studies into population properties, such as how likely was a LISA source's galactic progenitor to have a reversal in its magnetic field?

Current EM observatories have sensitivity to redshifts $z \lesssim 0.5$, which would include about half of the SMBH binary population in Fig.~\ref{F:Corner}, but detailed EM observations are only possible for $z \lesssim 0.01$. 
The future SKA facility will see to $z \lesssim 3$ and eliminate long-standing challenges to measuring magnetic fields, \cite{Gaensler2004} but will only be able to resolve large-scale magnetic field patterns out to distances of about $z < 0.1$ ($100$ Mpc). 
Meanwhile, these distances are smaller than a typical LISA source's distance, as indicated in the population in Fig.~\ref{F:Corner}, implying that LISA will open a new window into the study of magnetic fields to the distant and early Universe. 
As the distant LISA sources would appear as point sources for SKA, using the maxima of $|\mathbf{B}|$, $PI$, and $RM$ is a reasonable choice when probing galactic magnetic field populations. 
Thus LISA will revolutionize the study of galactic magnetic fields by revealing higher redshift galaxies and enabling multi-messenger constraints. 

Utilizing EM and GW observations from the same galaxies will allow for detailed analysis of magnetic field structure and its relation to central SMBHs, and correlating EM and GW observations of subsets of galaxies within a given redshift bin will provide population analyses of galactic magnetism as a function of redshift when applied to the cosmic horizon of LISA. 
A joint population analysis would need to account for the fact that EM data comes from an early source phase but only for the very local Universe and GW data comes from a later source phase but only for farther sources, resulting in sparse data and an interesting data analysis problem. 
SKA will also provide relief in studies of how $RM$ evolves with redshift which will be crucial for galaxies with weak polarization \cite{Gaensler2004}. A joint population analysis would need to consider contributions to $RM$ from sources along the line of sight such as the intergalactic medium and the Milky Way itself. 
For example, the intergalactic medium with a cosmologically aligned magnetic field $B = 10^{-9}$ G \cite{Grasso2001} can give $10 \lesssim RM \lesssim 1000$ rad/m$^2$ for redshifts $0.5 \lesssim z \lesssim 2$ which can be comparable to that of the source galaxy. 
As LISA will exist on an Earth-like orbit, it will also be important to account for the contribution due to the Milky Way whose magnetic fields are already known to be relevant for LISA data analysis \cite{Tartaglia2021}. 

Another area of significant overlap concerns the nanoHz GW detectors known as pulsar timing arrays (PTAs) that probe the stochastic GW background that can arise from populations of merging SMBHs \cite{NANOGravGWBAgazieEtAl:2023,ETPAGWBAntoniadisEtAl:2023,PPTAGWReardonEtAl:2023,CPTAGWBXuEtAl:2023,MPTAGWMilesEtal:2025} and possibly the individual mergers of very massive SMBHs \cite{Ellis2023}. 
The SMBH binaries in the population in Fig.~\ref{F:Corner} of Section~\ref{sec:LISA} finish the circumbinary disk evolutionary phase, which is where we compute the galactic magnetic field, with binary orbital frequencies $\sim 10^{-7}$ to $10^{-9}$ Hz placing their GW emission in the PTA regime. 
If LISA detects a population of BBH mergers, we can predict magnetic fields of their progenitor host galaxies to inform PTA searches for continuous waves from individually resolvable mergers. Similarly, EM measurements of galactic magnetic fields provide prior information on the host galaxy properties for such PTA searches. As SKA will revolutionize the study of both galactic magnetic fields and PTAs, we can combine these constraints for multi-messenger studies \cite{DeRosa2019,Charisi2022} of the galactic magnetic field population, implying great multi-band synergy as well with LISA. 
Ultimately, these multi-messenger and multi-band approaches can be combined for independent constraints on galactic magnetic fields and on the entire evolution of the SMBH binary population. 
    
Further future work includes multi-messenger studies of:
(i) galaxy morphology and its correlations with the magnetic field structure which may depend on possible selection effects of LISA detecting SMBH mergers in certain galaxy types; 
(ii) sources of the magnetic field itself, such as thin disk, gaseous halo, dark matter, etc, which might be challenging to disentangle; 
(iii) the hierarchical formation and cosmological evolution of galaxy clusters \cite{Durrer2013} and connections with the magnetic field of the intracluster medium \citep[e.g.,][]{Xu2010}; 
(iv) the early Universe where intergalactic magnetic fields can emerge from e.g. first order phase transitions \cite{Ellis2019}; and,
(v) comparison of EM-observed $RM$ maps with synthetic $RM$ maps \cite{Pakmor2018,Reissl2023} inferred from LISA data analysis of SMBH mergers.

\section{Conclusions}
\label{sec:ConcDisc}

This work presents a new way of probing cosmic magnetism with GW signals; it is the first study presenting synergies between radio observations of large-scale magnetic fields 
and LISA shedding light on probing the magnetic fields of galaxies and their connection to their SMBHs. 
Our main conclusions are that: 
\begin{enumerate}
    \item LISA will open a new window into the study of galactic magnetic fields, requiring detailed and accurate models of the astrophyscal and cosmological evolution of the magnetic field and its relation to the SMBH binary in the galactic core. 
    
    \item Combining this new window with predictions for interesting properties of the magnetic field, such as field reversals, and the EM observables of the progenitor system of the SMBH binary  generates new multi-messenger strategies for joint constraints on magnetic field properties and evolution. 

    \item The LISA signal-to-noise ratio $\rho$ is correlated with the maximum magnetic field via their joint dependence on the SMBH binary total mass $M$, implying a deep connection between this main LISA target and galactic stellar and gas dynamics \cite{Sesana2014} bridged by the study of galactic magnetic fields. 

    \item The relative uncertainties on the magnetic field magnitude $B$ can be inferred from LISA measurements of the SMBH parameters, such as the total mass $M$ which are typically $\lesssim 10^{-2}\,{\rm M}_\odot$ for binaries with $M \approx 10^7-10^8\Msol$ \cite{Klein2016,Pratten2023}. 
\end{enumerate}
This multi-messenger view motivates new frameworks for the astrophysical study of galactic magnetic fields, which must be formulated in a common language between LISA and galactic magnetic field observers and theorists \cite{Beck2019}. 
We emphasize that we do not claim that SMBHs or SMBH binaries cause galactic magnetic fields; rather, we propose that there are likely to be indirect astrophysical relations between them, and that these can be studied by LISA data analysis and source progenitor modeling. 

With numerous planned EM and GW detectors for the near future, and with ever expanding applications of multi-messenger astrophysics, our understanding of galactic magnetic fields is set to be revolutionized in the coming decades. This will necessitate accurate modeling frameworks and multi-disciplinary observational campaigns, implying also the potential emergence of entirely new astrophysical and cosmological fields of inquiry.

\acknowledgements
We thank Jayanne English, Krista Lynne Smith, Luke Zoltan Kelley, and Gautham Narayan for helpful discussions. N.S., S.S.H., and J.C.B. are supported by the Natural Sciences and Engineering Research Council of Canada (NSERC) through the Discovery Grants Program; N.S. and S.S.H are also supported through the NSERC Canada Research Chairs programs. 
Computations described in this paper were performed using the University of Manitoba’s Grex High Performance Computing (HPC) service, which provides an HPC service to the University’s research community. See \href{https://umanitoba.ca/information-services-technology/research-computing/um-high-performance-computing-system-grex}{here} for more details.
Besides the software tools cited in the main text, this work has made use of \textsc{corner}~\citep{corner}, \textsc{matplotlib}~\citep{matplotlib}, \textsc{numpy}~\citep{numpy}, a \textsc{scipy}~\citep{scipy}.

\bibliography{lisa-bfields}
\end{document}